\documentstyle[prb,aps,amsfonts,amssymb,multicol,epsf]{revtex}

\begin{document}
\draft

\title
{\bf Marginal pinning of vortices at high temperature}

\author{M.\ M\"uller$^{\,1,2}$,
  D.A.\ Gorokhov$^{\,1,3}$, and G.\ Blatter$^{\,1}$}

\address{$^{1\,}$Theoretische Physik,
  ETH-H\"onggerberg, CH-8093 Z\"urich, Switzerland}
\address{$^{2\,}$LPTMS, Universit\'e Paris Sud, B\^at. 100, F-91405 Orsay, France}
\address{$^{3\,}$Department of Physics,
  Harvard University, Cambridge, Massachusetts 02138}

\date{\today}
\maketitle

\begin{abstract}
We analyze the competition between thermal fluctuations and
pinning of vortices in bulk type II superconductors subject to
point-like disorder and derive an expression for the temperature
dependence of the pinning length $L_c(T)$ which separates different
types of single vortex wandering. Given a disorder potential with a basic
scale $\xi$ and a correlator $K_0(u) \sim K_0\, (u/\xi)^{-\beta}
\ln^\alpha (u/\xi)$ we determine the dependence of $L_c(T)$ on the
correlator range: correlators with $\beta >
2$ (short-range) and $\beta <2$ (long-range) lead to the known results $L_c(T) \sim L_c(0)
\exp[C T^3]$ and $L_c(T) \sim L_c(0) (C T)^{(4+\beta)/(2-\beta)}$, respectively.
Using functional renormalization group we show that for $\beta =2$
the result takes the interpolating form $L_c(T) \sim L_c(0) \exp[C
T^{3/(2+\alpha)}]$. Pinning of vortices in bulk type II
superconductors involves a long-range correlator with $\beta=2$,
$\alpha=1$ on intermediate scales $\xi<u<\lambda$, with $\xi$ and
$\lambda$ the coherence length and London penetration depth, hence
$L_c(T) \sim L_c(0) \exp[C T]$; at large distances $L_c(T)$
crosses over to the usual short-range behavior.
\end{abstract}


\begin{multicols}{2}

Many properties of type II superconductors derive from the
interaction between vortices and pinning
centers\cite{ScheidlNattermann,Blatter}. Impurities are
particularly important in situations where a transport current is
applied as they provide the necessary pinning force compensating
the Lorentz force acting on vortices\cite{Tinkham} and thus enable
a dissipation-free current flow. In this note we address specific
aspects of the high temperature pinning behavior of vortices
subject to weak point-like disorder. The results are of particular
relevance for the copper-oxide high $T_c$ superconductors which
can be operated at high temperatures and where such a pinning
landscape naturally derives from oxygen vacancies.

A single flux line in a disorder potential is described by the
partition function\cite{HalpinHealy}
\begin{eqnarray}
Z({\bf u},&L&)=\int_{({\bf 0},0)}^{({\bf u},L)}{\mathcal D}[{\bf
u}'(z)]
\nonumber\\
\label{partition} &
&\exp\biggl\{-\frac{1}{T}\int_0^L\,dz\biggl[\frac{c}{2}\Bigl(\frac{d{\bf
u}'}{dz}\Bigr)^2+V({\bf u}'(z),z]\biggr)\biggr\},
\end{eqnarray}
with ${(c/2)(\partial_{z}{\bf u})^{2}}$ the elastic energy and
$V({\bf u}, z)$ the disorder potential which might take negative
values. The disorder is chosen to be a Gaussian random variable
with zero mean and a correlator
\begin{equation}
\langle V({\bf u}, z)V({\bf 0}, 0)\rangle = K_{0}({\bf u})\delta
(z),
\end{equation}
where $\langle \dots \rangle$ denotes the average over disorder
realizations. The correlator $K_{0}({\bf u})$ decays on a length
$\xi$; in most applications the function $K_{0}({\bf u})$ is
assumed to be rapidly decaying, and at sufficiently high
temperatures the physical behavior is determined by the integral
$\Delta \equiv \int d^n u\thinspace K_{0}({\bf u})$ alone. In this
paper we draw attention to the situation in disordered type II
superconductors where the correlator $K_{0}({\bf u})$ describing
the potential landscape of vortices is long-ranged,
\begin{equation}
K_{0}({\bf u})\sim K_0\frac{\xi^2}{u^2} \ln\frac{u}{\xi},
\label{asymptotics}
\end{equation}
in the intermediate asymptotic regime $\xi < u < \lambda$ (here,
$\xi$ and $\lambda$ denote the coherence length and the London
penetration depth of the superconductor). The long-range tail
(\ref{asymptotics}) renders the correlator non-integrable and
modifies the pinning characteristics as compared to a short-range
correlated disorder landscape. This effect is particularly
pronounced for vortices in bulk high-$T_c$ superconductors where
the disorder becomes marginal at high temperatures and the ratio
$\lambda/\xi$ is large (below we describe the situation in an
isotropic material; effects of anisotropy can be accounted for
within the scaling approach\cite{BGL}).

The following discussion is not restricted to single vortex
pinning; rather we consider the latter as a specific realization
of the directed elastic string (or polymer-)
problem\cite{HalpinHealy} which describes such diverse physical
systems as domain walls in magnetic films\cite{Lemerle,Forgacs},
wetting  (in the plane)\cite{Wilkinson}, vortices\cite{Blatter} or
random polymers\cite{Kardar}. The numerous non-trivial features
that these systems have in common derive from an intricate
interplay of elasticity, disorder and thermal fluctuations. While
the elastic forces tend to stretch the string, the disorder
potential favors configurations deviating from a straight line in
order to take advantage of the potential valleys. Within the weak
collective pinning scenario\cite{Larkin} the elasticity dominates
on scales smaller than the cross-over scale $L_c$ and forces the
string to stay in the same valley, whereas on larger scales the
string effectively divides up into segments of size $L_c$ which
adjust independently to the disorder landscape. Increasing the
disorder strength decreases $L_c$, while thermal fluctuations tend
to smooth the disorder landscape implying an increase of $L_c(T)$
with temperature. Also, the typical barriers separating adjacent
valleys are reduced by thermal fluctuations leading to a peculiar
form of the creep-type dynamics under a small external
force\cite{GBM}. Below we first discuss some general properties of
vortex pinning in disordered type II superconductors and derive
the asymptotic form (\ref{asymptotics}) of the correlator. Second,
we calculate the pinning length $L_c(T)$ using the functional
renormalization group (FRG) approach\cite{DSFisher}.

The behavior of random directed polymers strongly depends on the
number $n$ of transverse motional degrees of freedom. E.g., the
(low temperature) roughness as characterized by the wandering
exponent $\zeta_n =\lim_{L\rightarrow \infty}{\partial \ln \langle
u^2(L) \rangle}/{\partial \ln L^2}$ decreases with $n$ (here, $L$
denotes the length of the polymer segment and ${\bf u}(L)$ is the
relative transverse displacement of its end points). Upon
increasing the temperature a phase transition (roughening
transition) is known to occur for $n\ge3$, see Ref.\
\onlinecite{Imbrie}, the high temperature phase being dominated by
thermal fluctuations on all length scales with a thermal roughness
exponent $\zeta_{\rm th}=1/2$. On the other hand, for $n=1,\,2$
the large scale behavior remains dominated by disorder with the
same exponent $\zeta_n$ as at low temperatures, however, going
beyond the so-called depinning temperature $T_{\rm dp}$, the
crossover scale $L_c(T)$ increases rapidly with temperature. In
the physically important case $n=2$ the crossover length $L_c(T)$
grows exponentially, $L_c(T)\sim L_c(0)\exp[C(T/T_{\rm dp})^\nu]$,
reflecting the fact that $n=2$ is the lower critical dimension of
the roughening transition. While $\nu=3$ is a well established
result describing the situation for a short-range disorder
potential\cite{Feigelman}, the exponent is modified by the
long-range tail of the potential correlator as it appears in the
vortex problem: we will show below that $\nu = 1$ (a similar
effect is found in the context of individual vortices pinned onto
columnar tracks\cite{Blatter}). We emphasize that the asymptotic
behavior (\ref{asymptotics}) does not influence the roughness of
the polymer on large scales, that is, the value of the wandering
exponent $\zeta_n$ remains unchanged. While for $n=2$ the
non-integrability of the function $K_{0}(u)$ with an asymptotic
decay slower than $1/[u^2\ln(u)]$, is sufficient to change the
value of the exponent $\nu$, the criterion on the asymptotics
$K_0(u)\sim K_0(\xi/u)^\beta$ to change the value of $\zeta_2$ has
been argued\cite{HalpinHealyPRL,BalentsFisher} to be
$\beta<3/\zeta_{2,{\rm sr}}-4$. Inserting the numerically known
value $\zeta_{2,{\rm sr}}\approx 5/8$ for the short-range
wandering exponent, we see that the $\zeta_2=\zeta_{2,{\rm sr}}$
is unaffected by the weak non-integrability ($\beta=2$) in the
vortex pinning problem.

We briefly derive the form of the long-range correlations
occurring in the problem of single vortex pinning in type II
superconductors\cite{Blatter}. The Ginzburg-Landau equation for
the macroscopic wave function $\Psi=\sqrt{\rho(R)}\exp(i\varphi)$
takes the form
\begin{equation}
   \xi^2 \Bigl(\nabla+\frac{2\pi i}{\Phi_0}{\bf A}\Bigr)^2\Psi
   + (1-|\Psi|^2)\Psi = 0,
   \label{GL}
\end{equation}
where ${\bf A}$ denotes the vector potential, $\Phi_0 = hc/2e$ is
the flux unit, and we have normalized $|\Psi|$ to unity in the
asymptotic regime. We concentrate on the vortex solution where
the phase turn in $\varphi$ by $2\pi$ drives a circular vortex
current. On scales $R < \lambda$ we can ignore transverse
screening and using $(\nabla\varphi)^2 = 1/R^2$, the real part of
(\ref{GL}) simplifies to $(-\xi^2/R^2+1-|\Psi|^2)\Psi = 0$, hence
\begin{equation}
   |\Psi|^2 (\xi \ll R < \lambda) \approx 1 - \xi^2/R^2;
   \label{al_decay}
\end{equation}
the circular vortex current produces an order parameter
suppression decaying only slowly at small distances $R < \lambda$
(transverse screening quenches the current flow beyond the
screening length $\lambda$ and the suppression of the order
parameter is exponentially small). The same result follows from a
variational Ansatz \cite{Schmid,Clem}, $\Psi=R/(R^2+2\xi^2)^{1/2}
\exp(i\varphi)$. In high temperature superconductors $\lambda$ is
much larger than $\xi$ and the slow decay of the order parameter
extends over a wide region.

To fix ideas, we consider disorder in the critical temperature (so
called $\delta T_c$-disorder\cite{Blatter}) described through
spatial variations $\delta\alpha({\bf r})$ in the Ginzburg-Landau
parameter $\alpha$ with correlations $\langle\delta\alpha({\bf
r})\delta\alpha({\bf r}')\rangle=\gamma\delta^3({\bf r}-{\bf r})$.
The pinning energy of a vortex aligned along the $z$-axis and with
coordinates $({\bf u}(z),z)$ is given by
\begin{equation}
  E_{\rm pin}({\bf u},z)=\int d^2R \,
  p(|{\bf u}-{\bf R}|)\delta\alpha({\bf R},z),
\end{equation}
with the vortex form factor $p(R)=1-|\Psi(R)|^2 = 2\xi^2/
(R^2+2\xi^2)$. Within the range $\xi \ll u <\lambda$ the pinning
energy correlator assumes the form
\begin{eqnarray}
   \label{vortexcorr}
   && K_0(u) \equiv \langle E_{\rm pin}({\bf u},z)E_{\rm pin}({\bf
   0},0)\rangle\\
   &&= \gamma \delta(z) \int d^2R \, p(|{\bf u}-{\bf R}|)p(R)
   \sim K_0 \delta (z) \frac{\xi^2}{u^2}\ln\frac{u}{\xi} \nonumber,
\end{eqnarray}
with $K_0=2 \pi |\Psi_0|^4 \gamma \xi^2$ (we have assumed
$|\Psi_0| = 1$ in the derivation above). The integral over this
correlator diverges logarithmically and is cut off only at
the large scale $\lambda$. This long-range feature will have an
important effect on the cross-over scale $L_c(T)$ at high
temperatures which we are now going to calculate.

We analyze the system described by the partition function
(\ref{partition}) with the help of the functional renormalization
group (FRG) approach\cite{DSFisher}. Applying momentum shell
renormalization to the replicated Hamiltonian leads to the
following system of one loop equations (see Ref.\
\onlinecite{Gorokhov}) for the renormalized correlator $K_l$ and
temperature $T_l$,
\begin{eqnarray}
  \label{RGK}
    \partial_l K_l(|{\bf u}|)
     &=&[3-(4+n)\zeta]\, K_l(|{\bf u}|)\\
     &+&\zeta{\bf \nabla}\cdot({\bf u}K_l(|{\bf u}|) )+
     {\tilde T}_lK^{\mu\mu}_l(|{\bf u}|)\nonumber\\
     &+&I\left[K_l^{\mu\nu}(|{\bf u}|)K_l^{\mu\nu}(|{\bf u}|)/2
     -K_l^{\mu\nu}(|{\bf u}|)K_l^{\mu\nu}(0)\right],\nonumber\\
  \label{RGT}
    \partial_l {\tilde T}_l&=&(1-2\zeta){\tilde T}_l,
\end{eqnarray}
where $I=1/(\pi c^2\Lambda^3)$, ${\tilde T}_l=T_l/(\pi c
\Lambda)$, and $\Lambda^{-1}$ denotes the short-scale cut-off of
the theory. In (\ref{RGK}) we limited ourselves to the two replica
correlator $K_l$, neglecting higher replica terms generated at
high temperatures during momentum shell integration, see Ref.\
\onlinecite{Gorokhov}; an analysis including these terms reveals
that their feedback to the flow of $K_l$ is of the same order as
the non-linear terms in (\ref{RGK}) and hence the qualitative
results obtained below remain valid.

At zero temperature the fourth derivative of the correlator
$K_l^{\rm\scriptscriptstyle (4)}(0)$ diverges at the finite scale
$l_c(T=0)\approx \ln \{9/ [(n+8)I K_0^{\rm \scriptscriptstyle
(4)}(0)]\}$, signalling the emergence of a non-analyticity in the
correlator at the origin and indicating that perturbation
theory in the disorder breaks down beyond this scale. Still, the
FRG-flow remains well-defined, and for the case of short-range
disorder the correlator can be shown to flow rapidly towards a
non-analytic disorder-dominated fixed point beyond $l_c(0)$. The
crossover scale\cite{Blatter}
$L_c(0)=\Lambda^{-1}e^{l_c(0)}\approx (c^2/K^{\rm
\scriptscriptstyle (4)}_0(0))^{1/3}$ is naturally interpreted as
the typical length of independently pinned vortex segments within
the weak collective pinning theory; the same result is obtained
from a simple scaling argument equating the elastic and disorder
energies.

At finite temperatures the correlator remains analytic due to the
thermal smearing introduced via the term ${\tilde T}
K_l^{\mu\mu}$. However, we can still identify the crossover scale
$l_c(\tilde{T})$ as the value of $l$ where the non-linear terms in
Eq.~(\ref{RGK}) become of the same order as the linear terms,
indicating that the disorder cannot any longer be dealt with
perturbatively; beyond this scale $K_l$ is again driven towards
the ($T=0$) strong coupling fixed point.

At small scales, the non-linear terms in the flow equation
(\ref{RGK}) for the correlator $K_l$ can be neglected.
Furthermore, since we want to study high temperatures, we choose
the roughness exponent to take the thermal value $\zeta =
\zeta_{\rm th} = 1/2$ which is convenient since the physics
appears more transparently in the sequel, in particular, the
temperature does not renormalize when $\zeta=\zeta_{\rm th}$ and
$\tilde{T}_l=\tilde{T}$ (however, note that from a mathematical
point of view the physical results below are independent of this
particular choice of $\zeta$). The linearized flow equation can be
solved explicitly \cite{GBM} with the result
\begin{eqnarray}
\label{solK}
  K_l(u)&=&e^{(1-n/2)l}\Bigl[\frac{1}{4\pi
  \tilde{T}(1-e^{-l})}\Bigr]^{n/2}\\
  &&\times\int{\rm d}^nu'\,
  \exp{\left[-\frac{({\bf u}-{\bf u}'e^{ -l/2})^2}
  {4\tilde{T}(1-e^{-l})}\right]}K_0(u').\nonumber
\end{eqnarray}

Let us analyze the situation for short- and long-range correlated
disorder separately. If the disorder correlator is short-ranged,
the total weight $\Delta \equiv \int d^n u' K_0(u')$ is finite and
the expression (\ref{solK}) can be simplified at high temperatures
$\tilde{T} \gg \tilde{T}_{\rm dp} \equiv \xi^2e^{-l_c(0)}$,
\begin{equation}
  \label{fst}
  K_l(u) = e^{(1-n/2)l} \Delta {\tilde T}^{-n/2}
  \exp[-u^2/4{\tilde T}],
\end{equation}
for $\max[1,\ln(\xi^2/\tilde T)] \ll l < l_c(0)$. Note that for
$n=1$ the correlator grows exponentially, reflecting the relevance
of the disorder, whereas in the marginal case $n=2$ the correlator
flows to a fixed point. In this case we can interpret the
linearized flow equation as the Fokker-Planck equation for the
probability distribution $K_l(u)$ of a particle in a harmonic
potential $V(u)=u^2/4$, where $l$ plays the role of time and
$\tilde T$ is the temperature. The above fixed point corresponds
to the stationary Gibbs-Boltzmann distribution which is rapidly
approached at high temperatures.

In the case of long-range correlations the integration kernel in
Eq.~(\ref{solK}) cuts the non-integrable tails of the correlator
at $u' \sim (e^l \tilde T)^{1/2}$ and we may estimate the
renormalized correlator at the origin as
\begin{equation}
   \label{snd}
   K_l(0) \sim e^{(1-n/2)l}(4{\tilde T})^{-n/2}
   \int_0^{(e^l\tilde T)^{1/2}}K_0(u) u^{n-1}\,du.
\end{equation}

\begin{figure}
   \centerline{\epsfxsize=7.5cm \epsfbox{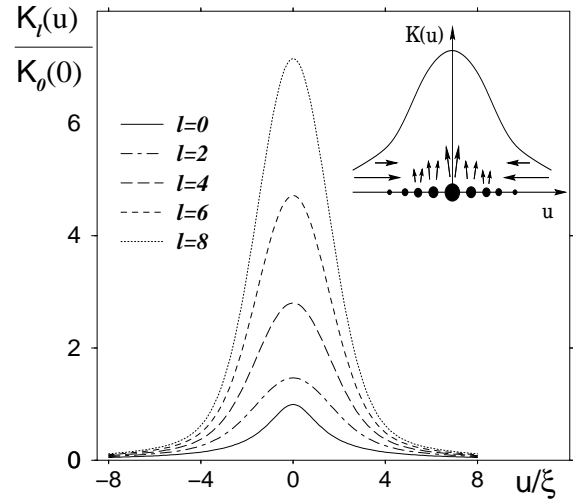}}
   \narrowtext\vspace{2mm}
            \caption{
The high temperature (${\tilde T}=\xi^2$) diffusive flow of the
correlator $K_l(u)$ evolving under the linearized FRG for $n=2$.
We choose a bare correlator $K_0(u)=\{1+\ln[1+(u/\xi)^2]/2\}
/[1+(u/\xi)^2]$ with non-integrable logarithmic tails as in the
vortex problem. After a crossover around $l=1$ the correlator
$K_l(0)$ increases quadratically with $l$ as expected from
Eq.~(13). The inset schematically illustrates the interplay of the
diffusive inflow of ``particles'' (horizontal arrows) and the
``particle'' sources originating from the quadratic one-loop terms
(solid dots). In the short-range case the first mechanism is
absent and $K_l$ grows due to the non-linear terms only, while in
the strongly long-range case ($\beta<2$), the non-linear terms
can be neglected and $K_l$ grows due to the inflow from the tails.
The vortex pinning problem is intermediate with a crossover at a
scale $l_{\rm cr}(\tilde{T}) < l_c(\tilde{T})$ from diffusion
driven growth to source driven growth. }
   \label{F1}
\end{figure}

In the case $n=2$ this expression can be understood in terms of the
diffusive motion of ``particles'' that are initially distributed
with a density proportional to $K_0(u)$. Their total number is
infinite due to the non-integrable tails of $K_0(u)$
(see Eq.~(\ref{vortexcorr})), and a growing number of ``particles''
from further and further away will eventually accumulate at the
origin, continuously increasing $K_l(0)$ (see Fig.~\ref{F1}).
Thermal equilibrium is never reached in this situation.

In the following we first discuss the situation where the
linearized flow gives rise to an exponential growth of the
renormalized correlator at the origin. This is the case for $n=1$,
or for $n=2$ if the initial correlator decays asymptotically as
$K_0(u)\sim K_0(\xi/u)^{\beta}$ with an exponent $\beta<2$. Under
these conditions, the contribution of the non-linear terms to the
growth of $K_l$ can be neglected, and the crossover scale
$l_c(\tilde{T})$ is found from comparing linear and quadratic
terms in the flow equation, i.e.,
\begin{equation}
   \label{crossover} I K^{\prime \prime}_{l_c(\tilde{T})}(0)
   \sim {\tilde T}.
\end{equation}

The second derivative of $K_l$ can be estimated from
Eq.~(\ref{solK}), $K^{\prime \prime}_l(0)\approx K_l(0)/{\tilde
T}$. Using expression~(\ref{snd}) in the above crossover condition
we find
\begin{equation}
   \label{Lc_lr}
   L_c(T>T_{\rm dp}) \sim L_c(0)
   \Bigl(\frac{T}{T_{\rm dp}}\Bigr)^{(4+\beta)/(2-\beta)}
\end{equation}
with the depinning temperature
\begin{equation}
   \label{T_dp}
   T_{\rm dp} \approx c \frac{\xi^2}{L_c(0)}.
\end{equation}
A short-range correlator with $n=1$ requires $\beta = 1$ and one
recovers\cite{VinokurIoffe} the well-known dependence $L_{c}(T)
\sim L_c(0) (T/T_{\rm dp})^{5}$. The result (\ref{Lc_lr}) can also
be derived from simple scaling arguments, c.f., Refs.\
\onlinecite{Blatter,GBM}.

The marginal case $n=2$ is more subtle if the correlator is
short-ranged or only weakly non-integrable with an asymptotic
behavior $K_0(u) \sim K_0\,(\xi/u)^{2}\ln^{\alpha}(u/\xi)$, where
$\alpha>-1$. In the short-range case the ``Boltzmannian'' fixed
point (\ref{fst}) is unstable only due to the quadratic loop
corrections in the FRG. They can be considered as sources of
``particles'' in the picture introduced above (see Fig.~\ref{F1}).
Well below the crossover $l_c(\tilde{T})$, these new particles
will quickly thermalize, and the shape of the distribution will
remain Boltzmannian, whereas its total weight will grow.
Qualitatively, we obtain $l_c(\tilde{T})$ as follows: From Eqs.\
(\ref{solK}) and (\ref{fst}) $K_l$ is seen to vary on a typical
scale $\tilde T^{1/2}$ thus the quadratic terms on the right-hand
side of Eq.~(\ref{RGK}) can be estimated by $AI[K_l(0)/\tilde
T]^2$ for $u \leq \tilde T^{1/2}$ with $A$ a numerical prefactor.
The amplitude of the correlator is determined from the flow
equation
\begin{equation}
    \label{approxflow} \partial_l K_l(0)
    \simeq A\frac{I}{\tilde T^2}K_l^2(0).
\end{equation}
Integrating over the interval ${\rm max}[1,\ln(\xi^2/\tilde{T})]
<l<l_c(\tilde{T})$ and using the crossover condition 
(\ref{crossover}) as well as the initial shape Eq.~(\ref{fst}) 
one recovers the well-known result $L_c(T)\sim L_c(0)
\exp[C(T/T_{\rm dp})^{3}]$. Here and in the following we make use
of the fact that the details of the initial condition $K_0(u)$ are
quickly washed out, giving way to a correlator of Boltzmann shape
(\ref{fst}); the latter serves as our new initial condition. For the
short-range case an exact flow equation for the total weight can
be written down under the assumption that the Boltzmannian shape
of the correlator is preserved. This allows to determine the
constant $C=32/\pi$ exactly, see Ref.\ \onlinecite{Gorokhov}.

Let us then discuss the case of a weakly long-range correlator
with a tail $K_0(u) \sim K_0 (\xi/u)^2
\ln^\alpha(u/\xi)$. Note that now the total weight of the
correlator diverges and the linearized flow does not approach a
fixed point. However, the flow of the central part of the
correlator is still governed by the linear terms in the flow
equation (\ref{RGK}) which force the correlator to maintain an
almost Boltzmannian shape of width $(4\tilde T)^{1/2}$. We will
therefore assume that the flow in the region $u\leq\tilde T^{1/2}$
is captured by the growth of the amplitude $K_l(0)$, and that
second derivatives may be replaced up to numerical factors by
$1/\tilde T$. In the first stage of the FRG-flow, the growth of
$K_l(0)$ is dominated by the diffusive flow of particles to the
origin as described by (\ref{snd}). However, at an intermediate
scale $l_{\rm cr}(\tilde{T})<l_c(\tilde{T})$ the quadratic source
terms become non-negligible and finally dominate, the subsequent
flow being analogous to the short-range case. In order to
determine the scale $l_{\rm cr}(\tilde{T})$ we compare the rate of
growth due to diffusion, Eq.~(\ref{snd}), with the magnitude of
the quadratic terms in Eq.~(\ref{RGK}),
\begin{equation}
\label{lcrcond}
    \partial_l K_l^{\rm diff}(0)\stackrel{(\ref{snd})}{\sim}
   \frac{\bigl[K_0(u)u^2\bigr]_{u=(e^{l_{\rm cr}}\tilde T)^{1/2}}}
   {8\tilde{T}}
   \sim
   I\biggl(\frac{K_{l_{\rm cr}}(0)}{\tilde{T}}\biggr)^2.\nonumber
\end{equation}
Assuming the bare correlator to decay asymptotically
as $K_0(u) \sim K_0\, (\xi/u)^2 \ln^\alpha(u/\xi)$ and evaluating
the right-hand side of (\ref{lcrcond}) with the help of
Eq.~(\ref{snd}) we find that
\begin{equation}
   \label{lcr}
   l_{\rm cr}(\tilde{T})\sim
   \Big(\frac{\tilde{T}}{\tilde{T}_{\rm dp}}\Bigr)^{3/(2+\alpha)}.
\end{equation}
Note that we have to require that $\alpha>-1$, otherwise the inflow of ``particles `` will saturate already at small scales and the crossover condition (\ref{lcrcond}) cannot be applied.

The further flow will be dominated by the non-linear terms while
the linear terms still make sure that the overall shape of $K_l$
resembles a Boltzmannian of width $(4\tilde T)^{1/2}$. Only beyond
the crossover scale $l_c(\tilde{T})$ will the non-linear terms be
large enough to drive the correlator to a disorder dominated fixed
point. As in the short-range case, $K_l(0)$ satisfies the
approximate equation (\ref{approxflow}) in the region $l_{\rm
cr}(\tilde{T})<l<l_c(\tilde{T})$. Integrating the latter beyond
$l_{\rm cr}(\tilde{T})$ and using the condition (\ref{crossover})
we find
\begin{equation}
   \label{lcTlcr}
   l_c(\tilde{T})-l_{\rm cr}(\tilde{T})
   \sim \Big(\frac{\tilde{T}}{\tilde{T}_{\rm dp}}\Bigr)^{3/(2+\alpha)},
\end{equation}
and, recalling $l=\ln(\Lambda L)$,
\begin{equation}
   \label{Lc_lr2}
   L_c(T)\sim
   L_c(0)\exp\Bigl[C\Bigl(\frac{T}{T_{\rm dp}}\Bigr)^{3/(2+\alpha)}\Bigr]
\end{equation}
with an unknown numerical factor $C$; its determination as well as
the precise value of the prefactor is beyond the accuracy of the
present analysis. The short-range result with an exponent
proportional to $T^3$ is recovered for $\alpha=-1$ which
corresponds to the limiting case of an integrable correlator
($\alpha<-1$). Eq.~(\ref{Lc_lr2}) shows that at a given (high)
temperature the crossover scale $L_c(T)$ is the smaller the larger
is the range of the potential correlations, turning finally to a
power law (\ref{Lc_lr}) for strongly long-range correlators with
$\beta<2$.
Note that while both the long- and short-range results derive from
cutting the flow through the condition (\ref{crossover}), the discussion
of the intermediate range correlator with $\beta=2$ and $\alpha>-1$ involves the additional crossover length $l_{\rm cr}(\tilde{T})$ which derives from
the condition (\ref{lcrcond}). The latter condition then determines the
exponent $3/(2+\alpha)$ providing the smooth interpolation between
the long- and short-range results.  

Single vortex pinning is described by $\alpha=1$ which leads to an
exponent $\propto T$. However, in the above derivation we have not
taken into account that the logarithmic tails (\ref{vortexcorr})
only extend up to $\lambda$. The result (\ref{Lc_lr2}) only
applies if at $l_{\rm cr}(\tilde{T})$ the integral (\ref{solK}) is
still cut by the Boltzmann kernel such that the estimate
(\ref{snd}) remains valid. We may obtain an upper bound
$\tilde{T}_u$ on $\tilde{T}$ by requiring that $(e^{l_{\rm
cr}(\tilde{T}_u)}\tilde{T}_u)^{1/2}=\lambda$ in the crossover
(\ref{lcrcond}). At temperatures higher than $\tilde{T}_u$ a
crossover to the short-range case will take place. Finally, the
range of validity of (\ref{Lc_lr2}) is found to be
\begin{equation}
\label{rangeofval}
    T_{\rm dp}< T< T_u \sim
    T_{\rm dp} \ln^{\frac{2+\alpha}{3}}(\kappa) =
    T_{\rm dp} \ln(\kappa).
\end{equation}
Apart from the restriction on the temperature the magnetic field
$B$ has to be weak such that effects due to vortex-vortex
interactions can be ignored; the corresponding condition $L_c(T)^2
< \Phi_0/B$ derives from comparing tilt and shear
energies\cite{Blatter}.

In conclusion, we have determined the temperature dependence of
the collective pinning length $L_c(T)$ for a directed elastic
string subject to a long-range correlated disorder potential. In
the physically important case of single vortex pinning in
three-dimensional bulk material $L_c(T)$ exhibits an exponential
sensitivity to temperature: in the intermediate temperature range
$T_{\rm dp} < T < (\ln\kappa)\,T_{\rm dp}$ the logarithmic tails of
the correlator produce a simple exponential law $L_c(T) \sim
L_c(0) \exp[C(T/T_{\rm dp})]$, while for $T>(\ln\kappa)\,T_{\rm dp}$ the usual
short-range result $L_c(T) \sim L_c(0) \exp[C(T/T_{\rm dp})^3]$
holds.

\end{multicols}

\end{document}